\documentclass[10pt]{article}
\usepackage{amsmath}
\usepackage{psfig}

\newcommand{\bv}{ {\bf v} }

\newcommand{\bd}{ {\bf d} }
\newcommand{\bS}{ {\bf S} }
\newcommand{\bT}{ {\bf T} }

\newcommand{\br}{ {\bf r} }

\newcommand{\bq}{ {\bf q} }

\newcommand{\hq}{ \hat q }

\newcommand{\Id}{ {\bf I} }
\newcommand{\bg}{ {\bf g} }
\newcommand{\bbg}{ {\bf \bar g}}
\newcommand{\bH}{ {\bf H} }
\newcommand{\bbH}{ {\bf \bar H}}
\newcommand{\bc}{ {\bf c} }

\newcommand{\bA}{ {\bf A} }
\newcommand{\bB}{ {\bf B} }
\newcommand{\bU}{ {\bf U} }

\newcommand{\bPhi}{ \boldsymbol {\Phi} }
\newcommand{\bbPhi}{ \boldsymbol {\bar \Phi} }
\newcommand{\baPhi}{ {\bar \Phi} }
\newcommand{\bPsi}{ \boldsymbol {\Psi} }
\newcommand{\bbPsi}{ \boldsymbol {\bar \Psi} }
\newcommand{\baPsi}{ {\bar \Psi} }
\newcommand{\bgamma}{ \boldsymbol{\gamma} }
\newcommand{\bagamma}{ \bar \gamma }
\newcommand{\bbgamma}{ \boldsymbol{\bar \gamma} }

\newcommand{\fz}{ \frac{1}{z} }

\newcommand{\fd}{\footnotesize{d}}
\newcommand{\dtot}[1]{ \frac{\mbox{\fd}}{\mbox{\fd} #1}}

\newcommand{\sech}{\text{sech}}

\renewcommand{\Re}{\text{Re} \;}
\renewcommand{\Im}{\text{Im} \;}

\begin{document}

\title{On Discretizations of the Vector Nonlinear Schr\"odinger Equation } 
\author{
  Mark J. Ablowitz\\
  Department of Applied Mathematics\\
  University of Colorado-Boulder\\
  Boulder, Colorado, 80309, USA\\
  \\
  Yasuhiro Ohta \\
  Department of Applied Mathematics, \\
  Faculty of Engineering, Hiroshima University \\
  1-4-1 Kagamiyama, Higashi-Hiroshima 739-8527, Japan\\
  \\
  A. David Trubatch\\
  Department of Applied Mathematics\\
  University of Colorado-Boulder\\
  Boulder, Colorado, 80309, USA\\ }

\date{}
\maketitle


\begin{abstract}
  Two discretizations of the vector nonlinear Schr\"odinger (NLS) equation are
  studied. One of these discretizations, referred to as the symmetric
  system,
  is a natural vector extension of the scalar integrable discrete NLS
  equation.
  The other discretization, referred to as the asymmetric system, has an
  associated linear scattering pair. General formulae for soliton
  solutions of the asymmetric system are presented.
  Formulae for a constrained class of
  solutions of the symmetric system may be obtained. Numerical
  studies support the hypothesis that the symmetric
  system has general soliton solutions.
\end{abstract}


\section{Introduction}

In recent years there has been wide interest in the study of solitons and
integrable systems. Researchers have found that not only are continuous
systems (i.e. PDE's) integrable via the inverse scattering transform (IST)
but also that interesting classes of discrete systems
(semi-discrete as well as partial difference equation) are integrable
-- cf. Ablowitz and Segur \cite{AbSe81} for an early review). In this letter,
we discuss soliton solutions and the integrable nature of certain discrete
systems associated with the vector extensions of the nonlinear
Schr\"odinger (NLS) equation:

\begin{equation}
  i q_t = q_{xx} + 2q|q|^2. \label{nls}
\end{equation}

\noindent NLS is a centrally important and physically significant nonlinear
equation which possesses solitons and is integrable via IST \cite{ZaSh72}.
Furthermore, NLS arises in many
areas of physics, such as the evolution of small
amplitude slowly varying wave packets in: deep water, nonlinear optics and
plasma physics (see e.g. \cite{AbSe81}). 

In 1974, Manakov \cite{Ma74} showed that the vector NLS (VNLS),

\begin{equation}
  i \bq_t = \bq_{xx} + 2 \left\| \bq \right\|^2 \bq  \label{vnls}
\end{equation}

\noindent where $\bq$ is an $N$-component vector and $\left\| \cdot \right\|$
denotes the vector norm,
also possessed solitons and could be integrated via IST (actually, in
\cite{Ma74} only the case $N=2$ was studied in detail; however the extension to
the $N$-th order vector system is straightforward). The second order ($N=2$)
VNLS equation is relevant in the study of electromagnetic waves
in optical media in which the electric field
has two nontrivial components. In optical fibers, the components of
$\bq$ in eq. (\ref{vnls}) correspond
to components of the electric field transverse to the direction of
wave propagation. These components of the transverse field compose a basis of
the polarization states. Although, in optical fibers, the equations governing
the field are in general a non-integrable variation  of vector
NLS \cite{Me87}, there are circumstances in which eq. (\ref{vnls}) is the
appropriate model \cite{EvMo92,Me89}.

In 1976, Ablowitz and Ladik \cite{AbLa76b} found that the following discrete
system is integrable via IST:

\begin{equation}
  i \dtot{t}q_n = \frac{1}{h^2} (q_{n-1} - 2 q_n +q_{n+1}) + 
    |q_n|^2 (q_{n+1}+q_{n-1})   \label{idnls}
\end{equation}

\noindent The continuum ($h \rightarrow 0$) limit of this discrete system
(\ref{idnls}) is NLS (\ref{nls}) hence this system can be referred to as 
integrable discrete NLS (IDNLS). The discrete system itself is also useful in
physical applications (see e.g. \cite{ClKi93,EiLo85,ItIz93,VaGa87}).

IDNLS provides an excellent numerical scheme to solve NLS (\ref{nls})
(cf. \cite{AbCl91}). Computations demonstrate
that, by simply replacing the nonlinear term in (\ref{idnls}) with
$2 |q_n|^2 q_n$, one obtains  a poor numerical scheme. In contrast, the
preservation of the integrable
structure in IDNLS (\ref{idnls}) plays a key role in its utility as a
numerical scheme. This points out that care must be taken in the choice of
discretization of the nonlinear terms for the purpose of numerical simulations.

It is natural to look for useful discretizations of VNLS (\ref{vnls}).
However, the special character of IDNLS as
compared to other discretizations of NLS, suggests that different
discretizations
of VNLS may have very different dynamics. Therefore, the choice of
discretization merits close analysis.

In this letter, we discuss two discretizations of VNLS which we refer to as
the symmetric and asymmetric discretizations.
These discretizations are as follows:

\begin{trivlist}

  \item Symmetric discretization:

        \begin{equation}
          i \dtot{t} \bq_n = \frac{1}{h^2} (\bq_{n-1} - 2 \bq_n + \bq_{n+1}) +
            \left\| \bq_n \right\|^2 (\bq_{n-1}+\bq_{n+1})  \label{sydvnls}
        \end{equation}

        \noindent where $\bq_n$ is an $N$-component vector.

  \item Asymmetric discretization:

        \begin{subequations}
        \begin{align}
          i \dtot{t} \bq_n &=
            \frac{1}{h^2} (\bq_{n-1} - 2 \bq_n + \bq_{n+1}) -
            (\br_n^T \bq_{n-1}) \bq_n - (\br_n^T \bq_n) \bq_{n+1}
            \label{asdvnlsq} \\
         -i \dtot{t} \br_n &=
            \frac{1}{h^2} (\br_{n-1} - 2 \br_n + \br_{n+1}) -
            (\br_n^T \bq_n) \br_{n-1} - (\br_{n+1}^T \bq_n ) \br_n
            \label{asdvnlsr}
        \end{align}
        \end{subequations}

        \noindent where, as before, $\bq_n$ is an $N$-component vector as is
        $\br_n$. The superscript $T$ denotes the transpose.

\end{trivlist}

Both systems (\ref{sydvnls}, \ref{asdvnlsq}-\ref{asdvnlsr})
reduce to VNLS (\ref{vnls}) in the continuum limit ($h \rightarrow 0$)
where in (\ref{asdvnlsq}-\ref{asdvnlsr}) we take $\br = -\bq^*$, where
$*$ denotes the complex conjugate,  after taking
the continuum limit. The asymmetric system (\ref{asdvnlsq}-\ref{asdvnlsr}),
however, does not admit the symmetry $\br_n = -\bq_n^*$ and hence remains
``asymmetric'' for $h$ finite.

These systems have a number of interesting properties. 
Both systems (\ref{sydvnls}, \ref{asdvnlsq}-\ref{asdvnlsr}) reduce to the
integrable discrete scalar case (\ref{idnls}),
just as VNLS (\ref{vnls}) reduces to NLS (\ref{nls}).
The asymmetric system (\ref{asdvnlsq}-\ref{asdvnlsr}) is associated with a
linear operator pair (see eq. \ref{asymsc}-\ref{asymtd}).
Although there is no known linear operator pair for the symmetric
system (\ref{sydvnls}),
under the reduction $\bq_n = e^{i \omega t} \bv_n$, where $\bv_n$ is
independent of $t$, this equation reduces to an
$N$-dimensional difference equation  which is known to be
integrable (cf. \cite{Su94}).

Both the symmetric (\ref{sydvnls}) and asymmetric
(\ref{asdvnlsq}-\ref{asdvnlsr}) systems possess a class of
soliton solutions. The asymmetric system has a class of soliton solutions
which correspond to all the soliton solutions of the continuum limit
NLS (\ref{vnls}). However, the exact multi-soliton solutions
found so far for the symmetric system (\ref{sydvnls}) only reduce to a
subset of those associated with eq. (\ref{vnls}). Since the known exact
soliton solutions of eq. (\ref{sydvnls}) are only a subset of those
known in the continuum limit (\ref{vnls}), we numerically examined the
interactions of solitary waves associated with the
symmetric system which lie outside this class.
The numerical evidence indicates that, in the symmetric system, the
solitary waves interact elastically and are therefore true
solitons. This  finding supports the possibility that the symmetric system
(\ref{sydvnls}) is indeed integrable.

%
%

\section{NLS, Integrable discrete NLS and Vector NLS}

We briefly review the relationship between NLS, IDNLS and Vector NLS in
order to clarify their relation to the symmetric and asymmetric
discretizations of vector NLS.

\subsection{NLS and IDNLS}

NLS (\ref{nls}) can be obtained as compatibility  condition of a
linear operator pair (e.g. a Lax pair). The linear pair can be used
to solve (\ref{nls}) by IST. The existence of an associated linear operator
pair and IST are hallmarks of integrability.
Therefore, in the effort to construct an integrable discretization
of NLS, it is  natural to look for a discrete version of the operator pair
and the compatibility condition. In fact, the integrable discretization of
NLS (\ref{idnls}) was obtained in exactly this manner \cite{AbLa76b}.

The extent to which the integrable structure of NLS is preserved under the
integrable discretization is significant: The IST for IDNLS is
analogous to the theory for NLS. In particular, the one-soliton solution of
IDNLS \eqref{idnls} given by

\begin{equation}
  q_n(t) = A e^{- i ( b h n - \omega t - \phi)} \sech(a h n - v t - \theta)
         = e^{i \phi} \hq_n(t)
  \label{discsol} 
\end{equation}

\noindent with

\begin{equation*}
       A = \frac{\sinh(a h)}{h}, \qquad
  \omega = \frac{2 (1 - \cosh(a h) \cos (b h))}{h^2},  \qquad
       v = 2 \frac{\sinh(a h) \sin(b h)}{h^2}
\end{equation*}

\noindent where $a$,$b$,$\theta$,$\phi$ are arbitrary parameters,
converges (as $h \rightarrow 0$) to the one-soliton solution of NLS with the
same four free parameters. More generally, solutions of IDNLS converge to the
solutions of NLS with an error of $O(h^2)$.

For both NLS and IDNLS, generic rapidly-decaying initial data resolves
itself into some number solitons plus radiation that vanishes as
$t \rightarrow \infty$ (in the sup norm).
Even though the solitons interact nonlinearly upon collision, it is well-known
that they retain their shape and speed after the collision. This can be
determined by comparing
the forward ($t \rightarrow +\infty$) and backward ($t \rightarrow -\infty$)
long-time limits of the solutions. In these long-time limits,
solitons traveling at different speeds are well-separated. Although
this collision is elastic, the position and overall complex phase of the
individual solitons are shifted between the forward and backwards long-time
limits. This elastic interaction of solitons up to a shift in phase
(position and overall complex phase) is typical of $(1+1)$-dimensional
integrable systems.

%
%
\subsection{VNLS} \label{subsection:vnls}

VNLS (\ref{vnls}) can be obtained by substituting
the vector $\bq$ for $q$ in 
NLS (\ref{nls}) and replacing complex conjugation with the Hermitian conjugate.
Similarly, the linear operator pair for VNLS can be obtained by appropriately
making the same substitution of scalars by vectors in the linear operator
pair for NLS (cf. Manakov \cite{Ma74}). 

When considered as a model for the propagation of electromagnetic waves in
optical fibers, the components of VNLS play the
role of a basis for the polarization vector. In the derivation of
VNLS by Manakov \cite{Ma74} the choice of basis is arbitrary. Therefore,
the vector system (\ref{vnls}) ought to be, and is,
invariant under a change of basis for the polarization. Mathematically, a
change of basis is obtained by multiplying the independent variable $\bq$ by
a unitary matrix, $\bU$. Therefore, the freedom in the choice of basis is
reflected in the fact that the vector system (\ref{vnls}) is invariant under
the transformation $\bq \rightarrow \bU \bq$. This symmetry is an important
feature in distinguishing among discrete versions of vector
NLS (notably, both the versions of discrete vector NLS discussed in this
letter (\ref{sydvnls},\ref{asdvnlsq}-\ref{asdvnlsr}) retain this symmetry). 

Under the reduction

\begin{equation}
  \bq = \bc \hq \label{reduce}
\end{equation}

\noindent where $\bc$ is a constant, $N$-component
vector such that $\left\| \bc\right\|^2 =1$ and $\hq$ is a scalar function of
$x$ and $t$, VNLS (\ref{vnls}) reduces to NLS (\ref{nls}). 
This is a manifestation of the fact that the vector system is a generalization
of NLS obtained by allowing the polarization to be non-constant.
When the solution of VNLS has a constant polarization, NLS is recovered.
As a consequence, any solution of scalar NLS has a corresponding family of
solutions of VNLS.
We call a solution of the form (\ref{reduce}) a {\em reduction solution}. 
Because the vector $\bc$ is arbitrary (up to the constraint that it is of
unit length), a reduction solution has correspondingly more parameters than
the solution of scalar NLS to which it reduces. We refer to the vector
$\bc$ in eq. (\ref{reduce}) as the {\em polarization} of the reduction solution.

The soliton solutions of vector NLS can be found by IST \cite{Ma74}.
The one-soliton solution of vector NLS is the reduction solution
(i.e of the form (\ref{reduce}) where $q_n$ is the one-soliton solution of
NLS. As in the scalar case, the IST \cite{Ma74} shows that solitons interact
elastically. By elastic, it is meant that the
solitons retain their shape and speed after interaction as shown by
comparing the forward and backward long-time limits.

For the vector system, care is needed:
the nature of the phase shift for vector solitons is somewhat different than
the scalar case. In addition to a shift in the center of the peak
(as in the scalar case),  vector solitons
also undergo a change in polarization upon collision.
That is, in the forward and backward long-time limit, the solution
asymptotically approaches a linear superposition of the individual solitons:

\begin{equation*}
  \bq \sim \sum_j \bq_j^{\pm} \qquad \text{as} \quad
    t \rightarrow \pm \infty
\end{equation*}

\noindent where $\bq_j^{\pm} = \bc_j^{\pm} \hq_j^{\pm}$  and, for each $j$, 
$\hq_j^+$ and $\hq_j^-$ are one-soliton solutions of NLS with the same
amplitude and speed. Comparison of the forward ($+$) and backward ($-$)
long-time limits shows that

\begin{equation*}
  \bc_j^- \neq \bc_j^+,
\end{equation*}

\noindent but

\begin{equation}
  ||\bc_j^+||^2 = ||\bc_j^-||^2 = 1. \label{vcons}
\end{equation}

\noindent A closed formula for this shift in the polarization can be
calculated by considering the
eigenfunction of the associated scattering problem \cite{Ma74}.

We refer to the squared modulus of a component of the  polarization vector 
as the {\em intensity} of that component of the polarization.
That is, if $c_j^{(\ell)}$ is the $\ell$-th component of
$\bc_j$, the polarization vector of the soliton $j$, then the intensity of
the of the $\ell$-th component of soliton $j$ is
$|c_j^{(\ell)}|^2$. The relation
(\ref{vcons}) implies that the total intensity of each
soliton is preserved, specifically the sum of the intensities is equal to one.
However, the distribution of intensity between the
components of the polarization of an individual soliton will not, in general,
be equal in the
forward and backwards long-time limits due to interaction with other solitons:
subject to the constraint of eq. (\ref{vcons}), in general,

\begin{equation*}
  |c_j^{(\ell)-}|^2 \neq |c_j^{(\ell)+}|^2.
\end{equation*}

\noindent This change in the distribution of intensity is a distinctive feature
of the vector system. There is no corresponding phenomenon in the soliton
interactions of the scalar equation (in both the scalar and vector equations,
the location of the peak of an individual soliton is shifted by the soliton
interaction).

Manakov \cite{Ma74} also observed the following special case for the
interaction of the solitons of VNLS: If, for every pair of solitons
$j$ and $k$, either

\begin{subequations}
\begin{gather}
  |\bc_j^- \cdot \bc_k^-| = 0 \label{vnls:orthogonal} \\
\intertext{or}
  |\bc_j^-\cdot \bc_k^-| = 1 \label{vnls:parallel}
\end{gather}
\end{subequations}

\noindent where $\cdot$ denotes
the dot product, then

\begin{equation}
  |\bc_j^- \cdot \bc_j^+| = 1 \label{vnls:noshift}
\end{equation}

\noindent for all $j$. As a consequence of \eqref{vnls:noshift},
$|c_j^{(\ell)-}|^2 = |c_j^{(\ell)+}|^2$ for all $j,\ell$.
Physically, this means that, in any polarization basis, the distribution of
intensity among the components of polarization for an individual soliton is
not changed by the soliton interaction.

%
%

\section{Asymmetric Discrete Vector NLS}

The preceding derivations of IDNLS and VNLS from NLS suggest
two methods to obtain an operator pair which has
a discrete form of VNLS as its compatibility condition: (i) discretize
the operator pair for vector NLS in a manner analogous to that used to
obtain the operator pair of IDNLS or (ii) appropriately
substitute vectors in to the operator pair for IDNLS. These approaches yield
the linear operator pair

\begin{subequations}
\begin{align}
  \bS_n &= \begin{pmatrix}
             z \Id_N & h \bq_n \\
             h \br_n^T & \fz 
           \end{pmatrix} \label{asymsc}\\
  \bT_n &= \begin{pmatrix}
             i  \bq_n \br_{n-1}^T - i \frac{1}{2 h^2} (z-\fz)^2 \Id_N &
            -i \frac{1}{h} \left(\fz \bq_n - z \bq_{n-1} \right) \\
             i \frac{1}{h} \left(\fz \br_n^T - z \br_{n-1}^T \right) &
            -i \br_n^T \bq_{n-1} + i \frac{1}{2 h^2} (z-\fz)^2
           \end{pmatrix} \label{asymtd}
\end{align}
\end{subequations}

\noindent where $\Id_N$ is the $N \times N$ identity matrix. The asymmetric
system (\ref{asdvnlsq}-\ref{asdvnlsr}) is the compatibility condition of this
pair.

Under the reduction

\begin{equation}
   \bq_n = \bc \hq_n, ~~ \br_n = \bd \hq_n^* \label{ared}
\end{equation}

\noindent where $\bc$ and $\bd$ are constant vectors such that
  $\bc \cdot \bd = 1$ and $\hq_n$ is a scalar function of $n$ and $t$,
the asymmetric
system (\ref{asdvnlsq}-\ref{asdvnlsr}) reduces to IDNLS \eqref{idnls}.
Solutions of this form \eqref{ared} are referred to
as {\em reduction} solutions. Hence, every solution of IDNLS generates a family
of reduction solutions of the asymmetric system.
We are particularly interested in reduction solutions where $\bd = -\bc^*$
because, in this case, $\br_n = -\bq_n^*$ and (\ref{asdvnlsq}-\ref{asdvnlsr})
reduces to a single equation.

\subsection{General Formula for Multi-soliton Solutions}

Multi-soliton solutions of the asymmetric system can be derived, without
the use of the IST machinery, by Hirota's method. First, make the
independent-variable transformations

\begin{equation*}
   \bq_n = \frac{\bg_n}{f_n}, \qquad \br_n = \frac{\bbg_n}{f_n}
\end{equation*}

\noindent where $\bg_n$ and $\bbg_n$ are vectors with $N$ components and
$f_n$ is a scalar. Then,  solutions of the system of bilinear equations

\begin{subequations}
\begin{gather}
  f_{n+1} f_{n-1} - f_n^2 = \bbg_n^T \bg_n \label{eqblbiga} \\
 i h^2 D_t f_n \cdot \bg_n = f_{n-1} \bg_{n+1} -2 f_n \bg_n +
    f_{n+1} \bg_{n-1} - \bH_n \bbg_n  \label{eqblbigb} \\
-i h^2D_t  f_n \cdot \bbg_n =  f_{n-1} \bbg_{n+1} -2 f_n \bbg_n +
    f_{n-1} \bbg_{n+1} - 
    \bbH_n \bg_n \label{eqblbigc} \\
  f_n \bH_{n+1} = h^2 \left( \bg_n \bg_{n+1}^T -
    \bg_{n+1} \bg_n^T \right) \label{eqblbigd} \\
  f_n \bbH_{n-1} = h^2 \left( \bbg_{n-1} \bbg_n^T -
    \bbg_n \bbg_{n-1}^T \right) \label{eqblbige}
\end{gather}
\end{subequations}

\noindent  where the $N \times N$ matrices $\bH_n$ and $\bbH_n$ are
auxiliary variables, are solutions of the
asymmetric system (\ref{asdvnlsq}-\ref{asdvnlsr}).

The $M$-soliton solutions for these bilinear equations
(\ref{eqblbiga}-\ref{eqblbige}) are
given by the following determinants:

\begin{subequations}
\begin{gather}
   f_n   = \begin{vmatrix} \bA_n & \Id \\ -\Id & \bB \end{vmatrix}, \qquad
   g_n^{(k)} = \begin{vmatrix}
                  \bA_n & \Id     & \Phi_n \\
                   -\Id & \bB     & 0 \\
                      0 & \bPsi_k & 0
               \end{vmatrix}, \qquad
   \bar{g}_n^{(k)} = \begin{vmatrix}
                        \bA_n    & \Id & 0 \\ 
                       -\Id      & \bB & \bbPsi_k \\
                        \bbPhi_n & 0   & 0
                     \end{vmatrix} \label{eqdetsolg}\\
   H_n^{(\ell,j)} = \begin{vmatrix}
                       \bA_n & \Id        & \bPhi_{n-1} & \bPhi_n \\
                      -\Id   & \bB        & 0           & 0 \\
                         0   & \bPsi_\ell & 0           & 0 \\
                         0   & \bPsi_j    & 0           & 0
                    \end{vmatrix}, \qquad
   \bar{H}_n^{(\ell,j)} = \begin{vmatrix}
                             \bA_n        & \Id & 0             & 0 \\
                            -\Id          & \bB & \bbPsi_{\ell} & \bbPsi_j \\
                             \bbPhi_n     & 0   & 0             & 0 \\
                             \bbPhi_{n+1} & 0   & 0             & 0
   \end{vmatrix} \label{eqdetsolh}
\end{gather}
\end{subequations}

\noindent where: $\Id$ is the $M \times M$ identity matrix;

\begin{trivlist}

  \item (i) $\bA_n$ and $\bB$ are $M \times M$ matrices defined by

    \begin{equation*}
      A_n^{(\ell,j)} = \frac{h}{e^{h(p_\ell+p_j^*)} -1}
        e^{\eta_{\ell,n}+\eta_{j,n}^*}, \qquad
      B^{(\ell,j)} = \frac{h}{e^{h(p_\ell^* + p_j)} -1}
        \bgamma_\ell^H \bgamma_j
     \end{equation*}

    \noindent with

    \begin{equation}
      \eta_{j,n} = p_j n h +\frac{i}{h^2} \left(2 - e^{h p_j} -
        e^{- h p_j} \right) t;
        \label{deta}
    \end{equation}

    \noindent the complex numbers $p_j=a_j - i b_j$, $a_j > 0$, determine
    the amplitude and speed of the $j$-th soliton; the $N$-component complex
    vectors
    $\bgamma_j$ and  $\bbgamma_j$  ($j=1, \ldots, M$)
    determine the polarizations and envelope
    phases of the solitons;

  \item (ii) $\bPhi_n$, $\bbPhi_n$ are, respectively, the $M$-component
    column and row vectors

    \begin{equation*}
      \Phi_n^{(\ell)} = e^{-h p_\ell} e^{\eta_{\ell,n}}, \qquad
      \baPhi_n^{(j)} = e^{\eta_{j,n}^*};  
    \end{equation*}

  \item (iii) $\bPsi_k$, $\bbPsi_k$ are, respectively, the  $M$-component row
    and column vectors,

    \begin{equation*}
      \Psi_k^{(j)}  = -\gamma_j^{(k)}, \qquad
      \baPsi_k^{(\ell)} = e^{- h p_\ell^*} \bagamma_\ell^{(k)}.
    \end{equation*}

\end{trivlist}

\subsection{The One-Soliton Solution}

From the above formulae (\ref{eqdetsolg}-\ref{eqdetsolh}), the one-soliton
($M$=1) solution is 

\begin{subequations}
\begin{align}
  \bq_n &=
    \frac{\bgamma_1}{\left( \bbgamma_1^T\bgamma_1 \right)^{\frac{1}{2}}}
      e^{-i h b_1} \frac{\sinh(a_1 h)}{h} e^{i \beta_{1,n}}
      \sech(\alpha_{1,n} + \delta_1)
    \label{asymoneq}\\
  \br_n &=
    -\frac{\bgamma_1}{\left( \bbgamma_1^T\bgamma_1 \right)^{\frac{1}{2}}}
      e^{-i h b_1} \frac{\sinh(a_1 h)}{h} e^{-i \beta_{1,n}}
      \sech(\alpha_{1,n} + \delta_1)
    \label{asymoner}
\end{align}
\end{subequations}

\noindent where

\begin{subequations}
\begin{align}
  \alpha_{1,n} &= \Re \eta_{1,n} =
    a_1 h n - 2 \dfrac{\sinh(a_1 h) \sin(b_1 h)}{h^2} t \label{alph} \\
  \beta_{1,n}  &= \Im \eta_{1,n} =
    -b_1 h n + \dfrac{2 (1 - \cosh(a_1 h) \cos (b_1 h))}{h^2} \label{bet}
\intertext{and}
      \delta_1 &=
        \log \frac{1}{2} \left\{ \bbgamma_1^T\bgamma_1 \right\} -
        \log \left\{ \frac{e^{2 a_1 h}-1}{h} \right\}. \notag
\end{align}
\end{subequations}

\noindent Note that to get a solution
(\ref{asymoneq}-\ref{asymoner}) with $\delta_1$ real we restrict our
attention to the case where
$\bbgamma_1^T \bgamma_1$ is real and positive. 
The one-soliton solution  (\ref{asymoneq}-\ref{asymoner})
is a reduction solution-- i. e. of the form \eqref{ared} --where

\begin{equation*}
  \bc = \dfrac{\bgamma_1}{\left( \bbgamma_1^T\bgamma_1 \right)^{\frac{1}{2}}}
    e^{-i h b_1}, \qquad
  \bd = -\dfrac{\bbgamma_1}{\left( \bbgamma_1^T\bgamma_1 \right)^{\frac{1}{2}}}
    e^{i h b_1}
\end{equation*}

\noindent and the scalar function $\hq_n$ is given by eq. \eqref{discsol}--
the one-soliton solution of IDNLS --where $a=a_1$, $b = b_1$ and
$\theta = - \delta_1$. In this solution, $\bd = - \bc^*$ if, and only if,
$\bbgamma_1 = \bgamma_1^*$.

%
%

\subsection{Two soliton Interaction}

To show that, for $M > 1$, the determinants (\ref{eqdetsolg}-\ref{eqdetsolh})
indeed give a multi-soliton solution, we consider the long-time limits.
In these limits, solitons moving at different speeds are separated.
For concreteness, let $M=2$ and

\begin{equation}
  \sinh(a_1 h) \sin(b_1 h) > \sinh(a_2 h) \sin(b_2 h) \label{diffvol}
\end{equation}

\noindent for the given $a_1$,$b_1$,$a_2$,$b_2$. The condition
\eqref{diffvol} ensures that each soliton travels with
a different speed (the analysis below holds, in general for $M$ solitons
as long as each has a different speed).
When the solitons travel at different speeds, we
determine the asymptotic form of an individual
soliton by taking long-time limits in a coordinate frame moving with
that soliton.

The limits

\begin{equation}
  n \sim 2 \dfrac{\sinh(a_j h) \sin(b_j h)}{a_j h^3} t, \qquad
  t \rightarrow \pm \infty \label{travel}
\end{equation}

\noindent are long-time limits in a coordinate frame moving with
soliton $j$. In the limit \eqref{travel} with $j=1$

\begin{equation}
  \Re \eta_{1,n} = const. \quad \text{and} \quad
  \Re \eta_{2,n} \rightarrow \pm\infty \qquad \text{as} \quad
  t \rightarrow \pm \infty
  \label{one:eta:limit}
\end{equation}

\noindent The substitution of \eqref{one:eta:limit} into
eq. (\ref{eqdetsolg}-\ref{eqdetsolh}), where $M=2$, yields

\begin{equation}
  \bq_n \rightarrow \bq_{1,n}^{\pm} \qquad \text{and} \qquad
  \br_n \rightarrow \br_{1,n}^{\pm} \qquad \text{as} \quad
  t \rightarrow \pm \infty, \label{one:asymptotic}
\end{equation}

\noindent which is the asymptotic form of soliton 1 in the forward ($+$) and
backward ($-$) long-time limits. The long-time limits for soliton 2 are
similar. The coordinate frame of soliton 2 is obtained by the
the limit \eqref{travel} with $j=2$ which yields 

\begin{equation}
  \Re \eta_{1,n} \rightarrow \mp \infty \quad \text{and} \quad
  \Re \eta_{2,n} = const \qquad \text{as} \quad
  t \rightarrow \pm \infty
  \label{two:eta:limit}
\end{equation}

\noindent Note the change in relative sign between $\eta_{1,n}$ and $t$
in the coordinate frame of soliton 2-- the slower soliton --in the long-time
limit. To obtain

\begin{equation}
  \bq_n \rightarrow \bq_{2,n}^{\pm} \qquad \text{and} \qquad
  \br_n \rightarrow \br_{2,n}^{\pm} \qquad \text{as} \quad
  t \rightarrow \pm \infty, \label{two:asymptotic}
\end{equation}

\noindent which is the asymptotic form of soliton 2 in the forward ($+$) and
backward ($-$) long-time limits, substitute \eqref{two:eta:limit}
into (\ref{eqdetsolg}-\ref{eqdetsolh}), where $M=2$. Combining
\eqref{one:asymptotic} and \eqref{two:asymptotic} gives

\begin{equation*}
  \bq_n \sim \bq_{1,n}^{\pm} + \bq_{2,n}^{\pm} \quad \text{and} \quad
  \br_n \sim \br_{1,n}^{\pm} + \br_{2,n}^{\pm} \qquad \text{as} \quad
  t \rightarrow \pm \infty.
\end{equation*}

\noindent The effect of interaction on the solitons is determined by comparing
$\bq_{j,n}^-$ and $\br_{j,n}^-$ with $\bq_{j,n}^+$ and $\br_{j,n}^+$ for
$j=1,2$. We give the formulae for the asymptotic forms of the solitons
below.  

First, we consider soliton 1. In the backward long-time limit,

\begin{equation*}
  \bq_{1,n}^- = \bc_1^- \hq_{1,n}^-, \qquad
  \br_{1,n}^- = \bd_1^- \hq_{1,n}^{-*}
\end{equation*}

\noindent where

\begin{subequations}
\begin{align}
  \bc_1^- &= \frac{\bgamma_1}
    {\left( \bbgamma_1^T\bgamma_1 \right)^{\frac{1}{2}}}e^{-i h b_1}
    \label{one:c:backward} \\
  \bd_1^- &= -\frac{\bbgamma_1}
    {\left( \bbgamma_1^T\bgamma_1 \right)^{\frac{1}{2}}} e^{i h b_1}
    \label{one:d:backward}
\end{align}
\end{subequations}

\noindent and $\hq_{1,n}^-$ is a one-soliton solution of IDNLS
(the $\hq$ in eq. \eqref{discsol}) with $a=a_1$, $b=b_1$ and

\begin{equation}
  \theta = -\delta_1^-
    = -\log \frac{1}{2} \left\{ \bbgamma_1^T\bgamma_1 \right\} +
    \log \left\{ \frac{e^{2 a_1 h}-1}{h} \right\}. 
\end{equation}

\noindent As in the one-soliton case, we require that $\bbgamma_1^T\bgamma_1$
is real and positive. The forward long-time for soliton 1 is

\begin{equation}
  \bq_{1,n}^+ =  \bc_1^+ \hq_{1,n}^+ \qquad
  \br_{1,n}^+ = \bd^+ \hq_{1,n}^{+*}
\end{equation}

\noindent where

\begin{subequations}
\begin{align}
  \bc_1^+ &= \frac{1}{\chi} (e^{-h p_1} \sigma_2 - \rho^*)
    \left( \sigma_2 (\bbgamma_2^T\bgamma_2) \bgamma_1 - 
    \rho^* (\bbgamma_2^T\bgamma_1) \bgamma_2 \right)
    \label{one:c:forward}\\
  \bd_1^+ &= -\frac{1}{\chi} e^{-h p_1^*} (\sigma_2 - \rho)
    \left( \sigma_2 (\bbgamma_2^T \bgamma_2) \bbgamma_1 - 
    \rho (\bgamma_2^T \bbgamma_1) \bbgamma_2 \right)
    \label{one:d:forward}
\end{align}
\end{subequations}

\noindent and $\hq_{1,n}^+$ is a one soliton solution of IDNLS with
$a=a_1$, $b=b_1$ and

\begin{equation*}
  \theta = -\delta_1^+ = -\log \chi + \log \sigma_2 +
    \log \left\{ \bbgamma_2^T  \bgamma_2 \right\}  
\end{equation*} 

\noindent also

\begin{gather*}  
  \chi = \left\{ \left( \sigma_1 \sigma_2 - |\rho|^2 \right)
    \left(\sigma_1 \sigma_2 (\bbgamma_1^T\bgamma_1)(\bbgamma_2^T \bgamma_2)
      - |\rho|^2 (\bbgamma_1^T\bgamma_2)(\bbgamma_2^T \bgamma_1) \right)
    \right\}^{\frac{1}{2}}
\end{gather*}

\noindent and we require that $\bgamma_j$,$\bbgamma_j$ for $j=1,2$
are such that
$\chi$ and $\bbgamma_2^T \bgamma_2$ are real and positive.
The constants $\sigma_1$, $\sigma_2$, $\rho$ are given by

\begin{equation*}
  \sigma_1 = \frac{h}{e^{h(p_1+p_1^*)}-1}, \quad
  \sigma_2 = \frac{h}{e^{h(p_2+p_2^*)}-1}, \quad
  \rho = \frac{h}{e^{h(p_1+p_2^*)}-1}.
\end{equation*}

\noindent Hence, the expression $\left( \sigma_1 \sigma_2 - |\rho|^2 \right)$
is real and positive. Note that, although
$\bc_1^- \neq \bc_1^+$, $\bd_1^- \neq \bd_1^+$,
and $\delta_1^- \neq \delta_1^+$ the
parameters $a_1$ and $b_1$ are the same in both long-time limits. Thus,
soliton 1 has the same form in both long-time limits but undergoes a phase
shift due to interaction with soliton 2.

The asymptotic limits of soliton 2 are similar to those for soliton 1 except
that the relative change in sign in \eqref{two:eta:limit} reverses the
calculations, that is,

\begin{equation*}
  \bq_{2,n}^- =  \bc_2^- \hq_{2,n}^- \qquad \text{and} \qquad
  \br_{2,n}^- =  \bd_2^- \hq_{2,n}^{-*}
\end{equation*}

\noindent where $\bc_2^-$ has the form \eqref{one:c:forward}  and
$\bd_2^-$ has the form \eqref{one:d:forward} with the indices $1$ and
$2$ exchanged. The scalar function $\hq_{2,n}^-$ is a one-soliton
solution of IDNLS with $a=a_2$, $b=n_2$ and $\theta=-\delta_2^-$ where
$\delta_2^-$ is equal to $\delta_1^+$ with the indices $1$ and $2$
exchanged. Similarly,

\begin{equation*}
  \bq_{2,n}^+ =  \bc_2^+ \hq_{2,n}^+ \qquad \text{and} \qquad
  \br_{2,n}^+ =  \bd_2^+ \hq_{2,n}^{+*}
\end{equation*}

\noindent where $\bc_2^+$ has the form \eqref{one:c:backward}  and
$\bd_2^+$ has the form \eqref{one:d:backward} with the indices $1$ and $2$
exchanged and $\hq_{2,n}^+$ is a one-soliton
solution of IDNLS with $a=a_2$, $b=n_2$ and $\theta=-\delta_2^+$ where
$\delta_2^+$ is equal to $\delta_1^-$ with the indices $1$ and $2$
exchanged.

In order to construct a solution such that 

\begin{equation*}
  \br_{j,n}^- = -\bq_{j,n}^{-*}
\end{equation*}

\noindent for $j=1,2$ it is necessary and sufficient that 

\begin{subequations}
\begin{align}
  \bbgamma_1 &= \bgamma_1^* \label{symm:one}\\
  \bbgamma_2 &= \dfrac{1}{ \left\| \bgamma_1 \right\|^2 }
    \begin{pmatrix}
      |\gamma_1^{(1)}|^2 + |e^{-h a_1} \gamma_1^{(2)}|^2 &
      (1-e^{-2 h a_1}) \gamma_1^{(1)*} \gamma_1^{(2)} \\
      (1-e^{-2 h a_1}) \gamma_1^{(1)} \gamma_1^{(2)^*} &
      \left |e^{-h a_1} \gamma_1^{(1)} \right|^2 +
        \left| \gamma_1^{(2)} \right|^2
    \end{pmatrix} \bgamma_2^*. \label{symm:two}
\end{align}
\end{subequations} 

\noindent For this choice, $\chi$ and
$\bbgamma_j^T \bgamma_j$ for $j=1,2$ are real and positive as we
required . However, if (\ref{symm:one}-\ref{symm:two}) hold then

\begin{equation}
  \br_{j,n}^+ \neq -\bq_{j,n}^{+*} \label{symm:forward}
\end{equation}

\noindent for $j=1,2$.
In fact, $\left\| \br_{j,n}^+ - (-\bq_{j,n}^{+*}) \right\| = O(h)$. 
An alternative is to set

\begin{equation}
  \bbgamma_1 = \bgamma_1^* \qquad \text{and}
  \qquad \bbgamma_2 = \bgamma_2^*.
  \label{simple:two}
\end{equation} 

\noindent Then, if \eqref{simple:two} holds,
$\chi$ and $\bbgamma_j^T \bgamma_j$-- for $j=1,2$ --are real and positive and

\begin{equation*}
  \br_{1,n}^- = -\bq_{1,n}^{-*}, \qquad
  \br_{2,n}^+ = -\bq_{2,n}^{+*}
\end{equation*}

\noindent but 

\begin{equation*}
  \| \br_{1,n}^+ - (-\bq_{1,n}^{+*}) \| = O(h) =
  \| \br_{2,n}^- - (-\bq_{2,n}^{-*}) \|.
\end{equation*}

\noindent In contrast, recall that
the solutions of the scalar IDNLS converge to solutions of the scalar NLS
with $O(h^2)$ convergence. This unavoidable asymmetry of the
system (\ref{asdvnlsq}-\ref{asdvnlsr}) makes it less desirable as a discrete
approximation of VNLS. Hence, we turn our attention to the symmetric system.

%
%

\section{Symmetric Discrete Vector NLS}

The symmetric system \eqref{sydvnls} is the natural vector generalization of
IDNLS \eqref{idnls}. We now describe some important symmetries of the 
symmetric system.
In analogy with the scalar case (\ref{idnls}), the symmetric
system  can be thought of as the
reduction of the system

\begin{subequations}
\begin{align}
  i \dtot{t} \bq_n &= \frac{1}{h^2} \left( \bq_{n-1} -2 \bq_n + \bq_{n+1}
                        \right) -
                       \br_n^T\bq_n \left( \bq_{n-1} + \bq_{n+1} \right)
    \label{sydvnlsq}\\
 -i \dtot{t} \br_n &= \frac{1}{h^2} \left( \br_{n-1} -2 \br_n + \br_{n+1}
                        \right) -
                      \bq_n^T \br_n \left( \br_{n-1} + \br_{n+1} \right)
    \label{sydvnlsr}    
\end{align}
\end{subequations}

\noindent obtained by letting $\br_n = - \bq_n^*$, which is the symmetry
that is broken by the asymmetric system.

Under the reduction

\begin{equation}
  \bq_n =\bc \hq_n \label{symm:reduce}
\end{equation}

\noindent where $\left\| \bc \right\|=1$, the symmetric system reduces to the
scalar IDNLS (\ref{idnls}). 
Furthermore, if $\bq_n$ satisfies the symmetric system, then so does
$\bU \bq_n$ where $\bU$ is a unitary matrix (this symmetry is a discrete form
of eq. \eqref{reduce}). Therefore, the symmetric
discretization retains the reductions and symmetries
of the asymmetric system and has the additional symmetry
$\br_n = - \bq_n^*$.

Despite the above-mentioned symmetries, there is,
to date, no
known associated linear operator pair for the symmetric system.
Without such a pair, the system (\ref{sydvnlsq}-\ref{sydvnlsr}) cannot be
solved by IST. The question remains, however, whether the symmetric system is
integrable. In the absence of IST, the existence of multi-soliton solutions
provides strong circumstantial evidence of integrability.

In order to determine whether there are solitary waves which interact
elastically-- i.e. solitons, we first identify a solitary-wave solution of
the symmetric system: a solution of the form \eqref{symm:reduce} where
the scalar $\hq_n$ is a one-soliton solution of IDNLS
(as in eq. \eqref{discsol}) is
a solitary-wave solution of the symmetric system. Note that
such a solution can have any polarization $\bc$ subject only
to the constraint that $\| \bc \|^2 = 1$. In order determine whether
these vector solitary waves interact exactly, we use both direct and numerical
methods.

\subsection{Soliton solutions by Hirota's Method}

Soliton solutions of the symmetric system
can be found by Hirota's method. For concreteness,
we consider the case where $\bq_n$ has two components ($N=2$). Under the
independent variable transformation,

\begin{equation*}
   \bq_n = \frac{\bg_n}{f_n}
\end{equation*}

\noindent where $\bg$ is a vector and  $f_n$ is a scalar, solutions
of the bilinear equations 

\begin{subequations}
\begin{gather}
   i h^2D_t f_n \cdot \bg_n = f_{n-1} \bg_{n+1} - 2 f_n \bg_n +
      f_{n+1} \bg_{n-1}
      \label{eqblsmallt} \\
   f_{n+1} f_{n-1} - f_n^2 =  h^2 ||\bg_n||^2 \label{eqblsmalln}
\end{gather}
\end{subequations}

\noindent are solutions of the symmetric system.

The following solution of the bilinear equations yields a two-soliton solution:

\begin{subequations}
\begin{align}
   f       &= 1 + e^{\eta_{1,n} + \eta_{1,n}^*} +
              e^{\eta_{2,n} + \eta_{2,n}^*} +
              |B_1|^2 e^{\eta_{1,n} + \eta_{1,n}^* + \eta_{2,n} + \eta_{2,n}^*}
              \label{fsymm}\\
   g^{(1)} &= \frac{1}{h}
             \left( e^{h p_1} - e^{- h p_1^*} \right) e^{\eta_{1,n}}
             \left( 1 + B_1  e^{\eta_{2,n} + \eta_{2,n}^*} \right)
             \\
   g^{(2)} &= \frac{1}{h}
             \left( e^{h p_2} - e^{- h p_2^*} \right) e^{\eta_{2,n}}
             \left( 1 + B_2  e^{\eta_{1,n} + \eta_{1,n}^*} \right)
             \label{gtwosymm}
\end{align}
\end{subequations}

\noindent where $\eta_{j,n}$ and $p_j$ are as in eq.
(\ref{deta}) and 

\begin{equation*}
  B_1 = \dfrac{(e^{h p_1}-e^{h p_2})(e^{h p_1}+e^{h p_2^*})}
              { (e^{h (p_1 + p_2)}+1)(e^{h (p_1 + p_2^*)}-1)}, \qquad
  B_2 = -\dfrac{(e^{h p_1}-e^{h p_2})(e^{h p_1^*}+e^{h p_2})}
               { (e^{h (p_1 + p_2)}+1)(e^{h (p_1^* + p_2)}-1)} .
\end{equation*}

In order to see that the above solution indeed gives a two-soliton solution,
consider the long-time limits. We calculate the long-time limits
of the Hirota form of the solution (\ref{fsymm}-\ref{gtwosymm}) by the same
approach as was used for the asymmetric case.
In the forward ($+$) and backward ($-$) long-time limits, the solution
a (\ref{fsymm}-\ref{gtwosymm}) asymptotically approaches the form

\begin{equation*}
   \bq_n \sim \bq_{1,n}^\pm + \bq_{2,n}^\pm
\end{equation*}

\noindent as $t \rightarrow \pm \infty$.

The long-time limits for soliton 1 are

\begin{subequations}
\begin{align}
  \bq_{1,n}^- &= \begin{pmatrix} e^{i \phi_1^-} \\ 0 \end{pmatrix}
    \frac{\sinh(a_1 h)}{h} e^{i \beta_{1,n}} \sech(\alpha_{1,n})
    \label{onemin}\\
  \bq_{1,n}^+ &= \begin{pmatrix} e^{\phi_1^+} \\ 0 \end{pmatrix}
    \frac{\sinh(a_1 h)}{h} e^{i \beta_{1,n}} \sech(\alpha_{1,n} + \log |B_1|)
    \label{oneplu}
\end{align}
\end{subequations}

\noindent where

\begin{equation*}
  \phi_1^- =  h b_1, \qquad \phi_1^+ = h b_1 + \arg B_1
\end{equation*}

\noindent and $\alpha_{1,n}$, $\beta_{1,n}$ are as in (\ref{alph}-\ref{bet}).
In both limits, this is a solution of the form
$\bq_{1,n}^{\pm} = \bc_1^{\pm} \hq_{1,n}$  where $\hq_{1,n}^{\pm}$
is a one-soliton solution of IDNLS (as in \eqref{discsol}).
Note that the polarization, $\bc_1^{\pm}$, is such that
$|c_1^{(1)\pm}|^2 = 1$ (and, therefore, $|c_1^{(2)\pm}|^2 = 0$ since
$\|\bc_1^{\pm}\|^2=1$).

The long-time limits for soliton 2 are

\begin{subequations}
\begin{align}
  \bq_{2,n}^- &= \begin{pmatrix} 0 \\ e^{i \phi_2^-} \end{pmatrix}
    \frac{\sinh(a_2 h)}{h}
    e^{i \beta_{2,n}} \sech(\alpha_{2,n} + \log |B_1|)
    \label{twomin}\\ 
  \bq_{2,n}^+ &= \begin{pmatrix} 0 \\ e^{i \phi_2^+} \end{pmatrix}
     \frac{\sinh(a_2 h)}{h}
    e^{i \beta_{2,n}} \sech(\alpha_{2,n})
    \label{twoplu}
\end{align}
\end{subequations}

\noindent where

\begin{equation*}
  \phi_2^- = h b_2 + \arg B_2, \qquad
  \phi_2^+ = h b_2
\end{equation*}

\noindent  and $\alpha_{2,n}$, $\beta_{2,n}$ are as in
(\ref{alph}-\ref{bet}) with $a_2$ and $b_2$ replacing $a_1$ and
$b_1$ respectively. In both
the forward and backward long-time limit, this is a solution of the form
$\bc_2^\pm \hq_{2,n}^\pm$ where $|c_2^{(2)\pm}|^2 = 1$ (and, therefore,
$|c_2^{(2)\pm}|^2 = 0$ since $\|\bc_2^{\pm}\|^2=1$).

The solution (\ref{onemin}-\ref{oneplu}, \ref{twomin}-\ref{twoplu}) is not
the most general two-soliton interaction because

\begin{equation}
  |\bc_1^- \cdot \bc_2^-| = |\bc_1^+ \cdot \bc_2^+| =  0.
  \label{orthogonal}
\end{equation}
 
\noindent The two-soliton solution
(\ref{onemin}-\ref{oneplu}, \ref{twomin}-\ref{twoplu}) can be
multiplied by a unitary matrix to obtain a two-soliton solution with any
$\bc_1^-$,$\bc_1^+$ such that $|\bc_1^- \cdot \bc_2^-|=0$ 
(under any such transformation, the condition $|\bc_1^+ \cdot \bc_2^+| =  0$
will still hold). Thus, these two-soliton solutions are a constrained
class of two-soliton solutions where \eqref{orthogonal} is the constraint.

More generally, Hirota's method can be used to derive solutions of
the symmetric system with more than two solitons. These solutions
can be represented as combinations of Pfaffians
(these formulae are quite technical; they are presented in a
separate publication \cite{Oh98}). The $M$-soliton solutions derived
in this manner are constrained such that either

\begin{subequations}
\begin{gather}
  |\bc_j^- \cdot \bc_k^-| = 1 \label{symm:parallel}\\
\intertext{or}
  |\bc_j^- \cdot \bc_k^-| = 0 \label{symm:orthogonal}\\  
\end{gather}
\end{subequations}

\noindent for all $j,k=1, \ldots, M$. Moreover, the above satisfy

\begin{equation}
  |\bc_j^- \cdot \bc_j^+|=1 \label{symm:noshift}
\end{equation}

\noindent for all $j=1, \ldots, M$. These multi-soliton
solutions are reminiscent of the special case of soliton interactions in the
PDE discussed at the end of section \ref{subsection:vnls}: conditions
(\ref{symm:parallel}-\ref{symm:orthogonal})  and eq. \eqref{symm:noshift}
for the discrete symmetric system are the counterparts
of (\ref{vnls:orthogonal}-\ref{vnls:parallel}) and eq. \eqref{vnls:noshift} 
in the PDE. 
To date, there is no known analytic formula for a general multi-soliton
solution of the symmetric system such that
$0 < |\bc_j^- \cdot \bc_k^-| < 1$ and $|\bc_j^- \cdot \bc_j^+| < 1$.

%
%

\subsection{Numerical Simulation of Soliton Interactions}

The individual solitary waves of the symmetric system may have any
polarization. In the PDE (\ref{vnls}) such solitary waves interact as
solitons. That is, there are vector multi-soliton solutions of VNLS
with asymptotic polarizations $\bc_j^{\pm}$ such that
$0 < |\bc_j^- \cdot \bc_k^-| < 1$ and $|\bc_j^- \cdot \bc_j^+| < 1$.
In the absence of analytical formulae in the discrete symmetric system
for these general multi-soliton interactions, we investigated the
collision of solitary waves by numerical simulation.
 
\begin{figure}[p]
  \centerline{
    \psfig{figure=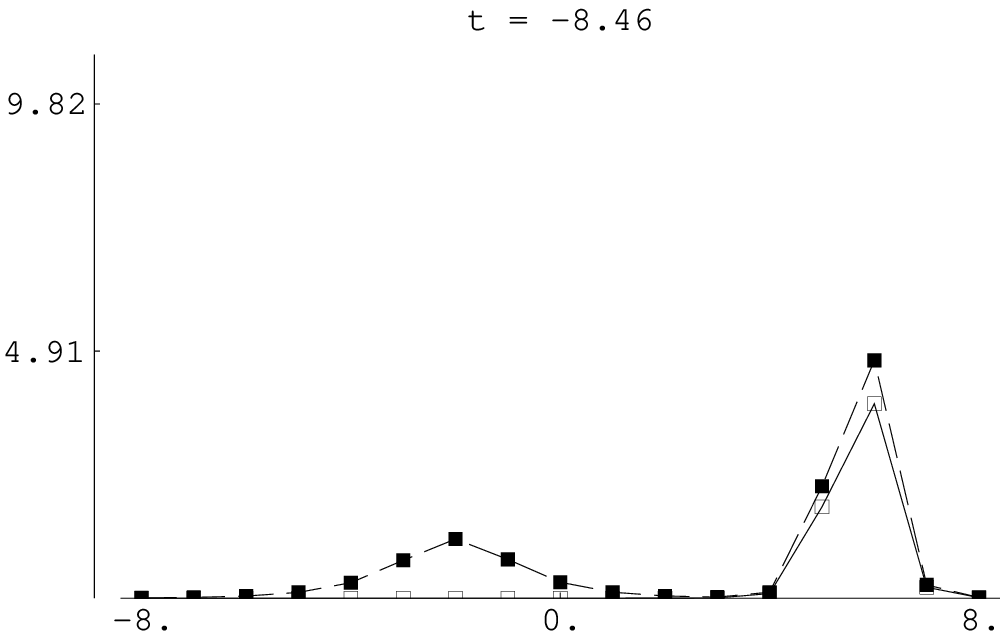,width=3in} \hfill
    \psfig{figure=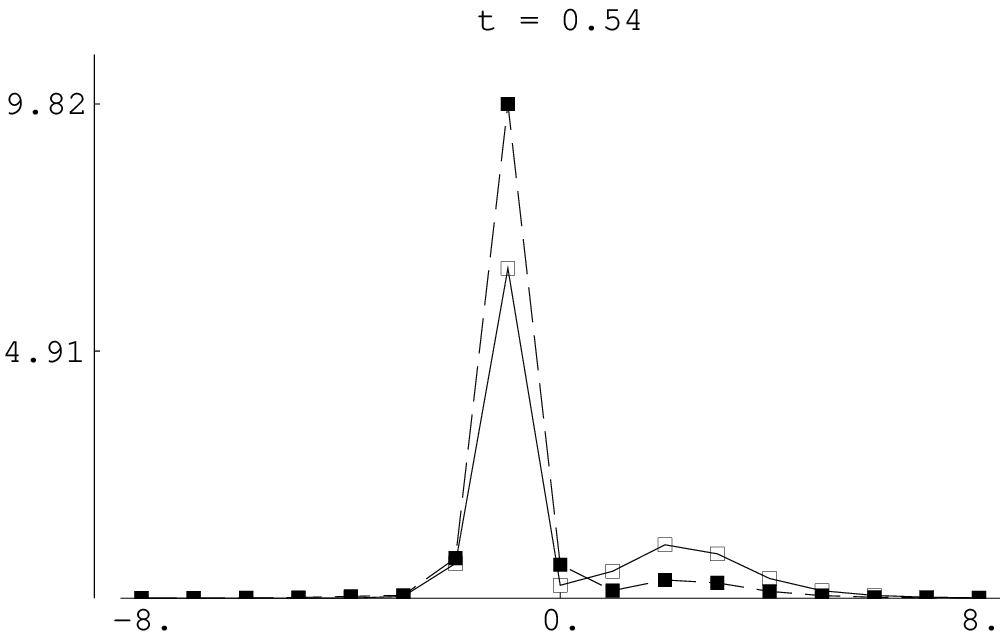,width=3in}} 
   \centerline{
     \psfig{figure=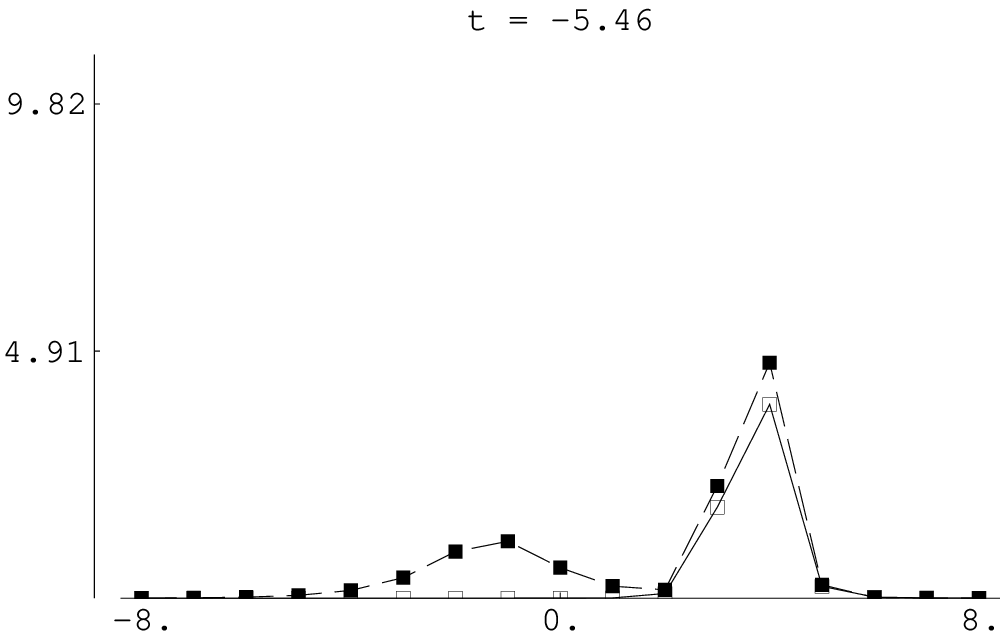,width=3in} \hfill
     \psfig{figure=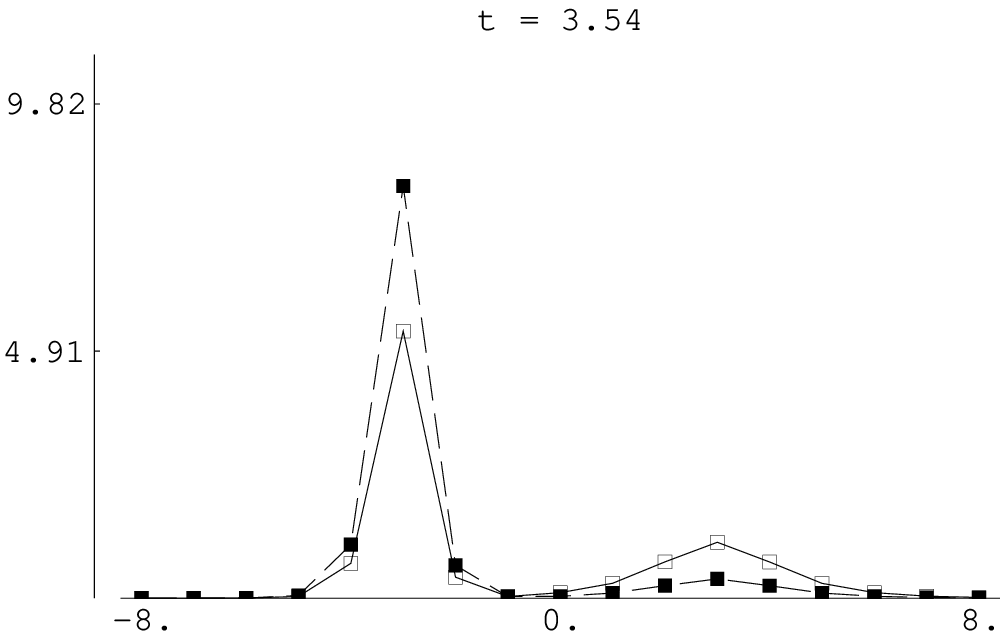,width=3in}} 
   \centerline{
     \psfig{figure=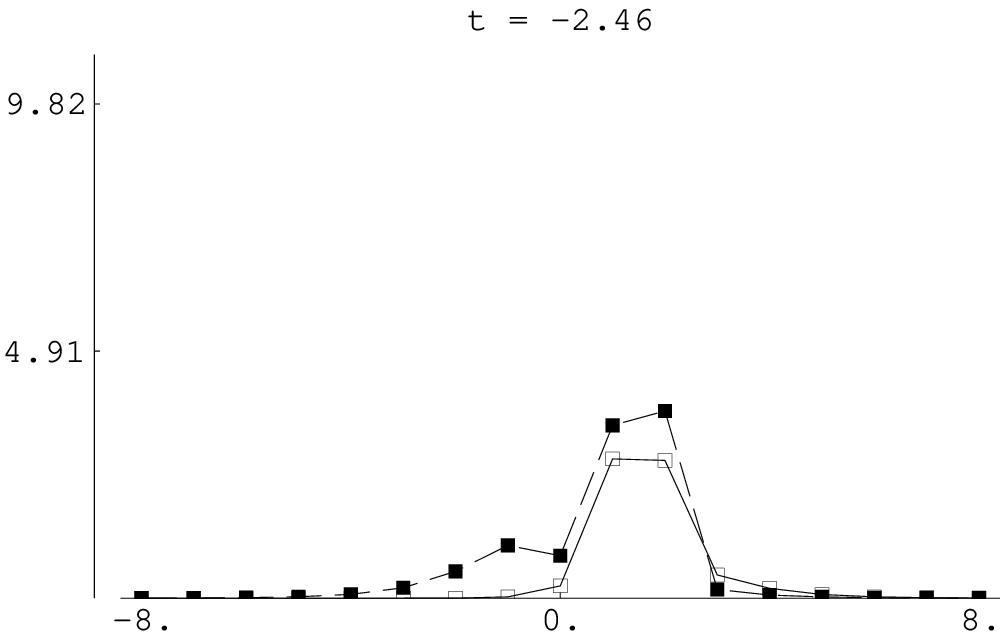,width=3in} \hfill
     \psfig{figure=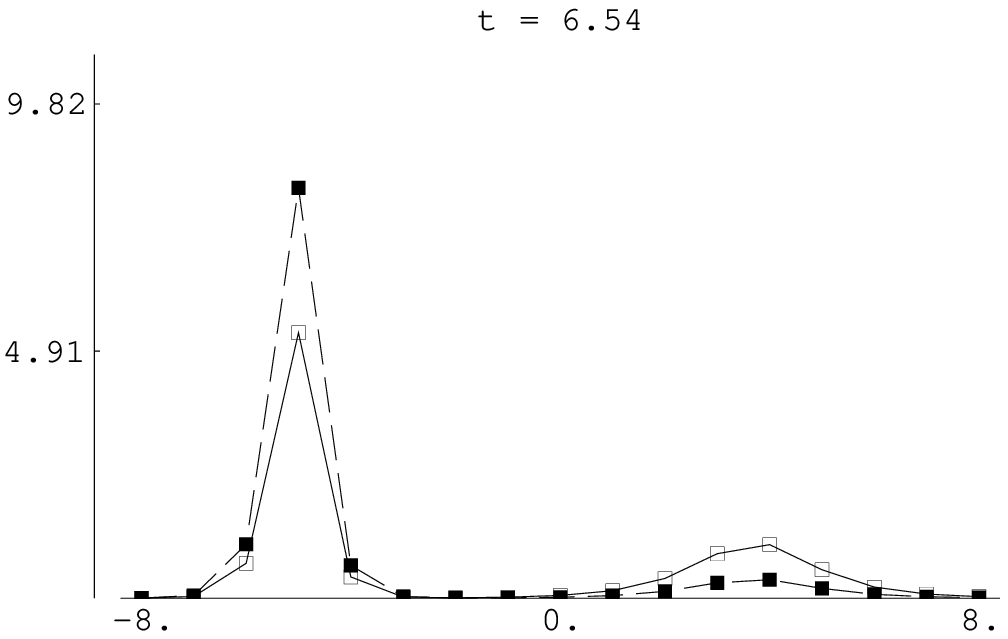,width=3in}} 
    \caption{
     Two-soliton interaction for the symmetric system.
     The filled boxes are $|q_n^{(1)}|$ and the open boxes are
     $|q_n^{(2)}|$. Increasing time is read down column-wise.
     Soliton 1 is on the left and Soliton 2 is on the right
     at $t=-8.46$. They are reversed at $t=6.54$.
     The soliton parameters are: $a_1 = 1$, $a_2 = 3$,
     $b_1 = .1$, $b_2 = - .1$.
     The polarization vectors are: $\bc_1^- = (1, 0)$,
     $\bc_2^- = (.60, .80)$ , at $t=-8.46$, and 
     $\bc_1^+ = (.33 e^{i .23 \pi}, .95 e^{i .12 \pi})$,
     $\bc_2^+ = (.84 e^{- i .02 \pi}, .54 e^{- i .01 \pi})$
     at $t=6.54$. This is a typical two-soliton interaction where
     $0 < |\bc_1^- \cdot \bc_2^-| = .6 < 1$,
     $|\bc_1^- \cdot \bc_1^+ | = .204 < 1$ and
     $|\bc_2^- \cdot \bc_2^+ | = .993 < 1$.
   }
   \label{fig:twosym}
\end{figure}

In the simulations, the initial conditions were taken to be of the form

\begin{equation*}
  \bq_n^- = \bq_{1,n}^- + \bq_{2,n}^-
\end{equation*}

\noindent where $\bq_{j,n}^-=\bc_j^- \hq_{j,n}^-$ and $\hq_{j,n}^-$ is
of the form \eqref{discsol}. Then, the symmetric system was integrated in
time by an adaptive Runge-Kutta-Merson routine (from the NAG library) until
the peaks were again well-separated (see Figure \ref{fig:twosym} for an
example). The separation of peaks in the initial and final conditions
makes these conditions comparable to, respectively, the backwards and
forwards long-time limits. The solitary-wave interactions simulated
in this manner comprised initial conditions in which
$0 < |\bc_j^- \cdot \bc_k^-| < 1$.
Visually, in Figure \ref{fig:twosym} (and in other simulations) the solitary
waves appear to interact without any radiation. We confirmed this finding by
quantitative comparison of the solitary waves at the final time and exact
solitary waves with the same height and speed parameters as the initial data.

The error at the final time was defined separately for each solitary
wave by

\begin{equation}
   \Delta_j = \frac{1}{A_j} \max_{n \in \Omega_j}
     \left\| \bq_{j,n}^f - \bq_{j,n}^+ \right\| \label{error}
\end{equation}

\noindent where:

\begin{trivlist}
  \item $\bq_{j,n}^f$ is the numerical data at the final time;
  \item $\bq_{j,n}^+$ is a solitary wave with the same amplitude and speed
    parameters-- $a_j$ and $b_j$ --as the initial data, $\bq_{j,n}^-$;
  \item $A_j= \tfrac{\sinh(a_j h)}{h}$ is the amplitude of the
    exact solitary wave with amplitude parameter $a_j$;
  \item $\Omega_j$ is the set of points containing the main contribution
    of the $j$-th solitary wave--
    i.e. $\Omega_j = \left\{ n : || \bq_{n,j}^f || > \epsilon \right\}$ \\
    for $\epsilon$ small compared to $\max_{j=1,2} A_j$.
\end{trivlist}
  
The resulting errors were small (see Table \ref{table:twosym} for an example).
Furthermore, when the user-supplied error bound in the adaptive integration
scheme was decreased, the errors, $\Delta_j$, decreased proportionally.
Therefore, the differences between the final wave forms obtained
by numerical simulation and exact solitary waves
are accounted for by errors in the time integration. These results,
strongly suggest that the solitary waves interact elastically-- 
i. e. the solitary waves are solitons.

In the numerical simulations described in Table \ref{table:twosym}, and
in other experiments we considered initial data  such that
$0< |\bc_1^- \cdot \bc_2^-| < 1$, the general case of the 
soliton interaction for which there is no known
analytical solution. In the PDE (\ref{vnls}), such conditions result in the
shift of the polarizations of the individual vector
solitons-- i.e. $|\bc_j^- \cdot \bc_j^+| < 1$, $j=1,2$. The
simulations described in Table \ref{table:twosym}, consistent with other
numerical experiments, show the same distinctive vector soliton behavior for
the  discrete symmetric system.
  
The mechanism of the more general elastic soliton interactions observed
in the symmetric system remains to be explained
analytically. More generally, proof that the symmetric system
is integrable remains as an important open problem.

\section*{Acknowledgments}

This effort was sponsored in part by the Air Force Office of Scientific
Research,  Air Force Materials Command, USAF, under grant number
F49620-97-1-0017, by
the Office of Naval Research, USN, under grant number N00014-94-1-0915 and
the National Science Foundation under grant number DMS-9703850.
The US Government is authorized to reproduce
and distribute reprints for governmental purposes notwithstanding any
copyright notation thereon. 
The views and conclusions contained herein 
are those of the authors and should not be interpreted as necessarily
representing the official policies or endorsements, either expressed
or implied, of the Air Force Office of Scientific Research or the US
Government. One of the authors (YO) is grateful to Dr. S. Tsujimoto
for valuable discussions and also acknowledges the financial support
of the Japan Ministry of Education through the Foreign Study Program.

\bibliographystyle{unsrt}
\bibliography{vnls}

\begin{thebibliography}{10}

\bibitem{AbSe81}
M.~J. Ablowitz and H.~Segur.
\newblock {\em Solitons and the Inverse Scattering Transfom}.
\newblock Number~4 in SIAM Studies in Applied Mathematics. SIAM, 1981.

\bibitem{ZaSh72}
V.~E. Zakharov and A.~B. Shabat.
\newblock Exact theory of two-dimensional self-focusing and one-dimensional
  self-modulation of waves in nonlinear media.
\newblock {\em Soviet Physics JETP}, 34:62--69, 1972.

\bibitem{Ma74}
S.~V. Manakov.
\newblock On the theory of two-dimensional stationary self-focusing of
  electromagnetic waves.
\newblock {\em Soviet Physics JETP}, 38(2):248--253, 1974.

\bibitem{Me87}
C.~R. Menyuk.
\newblock Nonlinear pulse propagation in birefringent optical fibers.
\newblock {\em IEEE Journal of Quantum Electronics}, QE-23(2):174--176,
  February 1987.

\bibitem{EvMo92}
S.~G. Evangelides, L.~F. Mollenauer, J.~P. Gordon, and N.~S. Bergano.
\newblock Polarization multiplexing with solitons.
\newblock {\em Journal of Lightwave Technology}, 10(1):28--35, January 1992.

\bibitem{Me89}
C.~R. Menyuk.
\newblock Pulse propagation in an elliptically birefringent {K}err medium.
\newblock {\em IEEE Journal of Quantum Electronics}, 25(12):2674--2682,
  December 1989.

\bibitem{AbLa76b}
M.~J. Ablowitz and J.~F. Ladik.
\newblock A nonlinear difference scheme and inverse scattering.
\newblock {\em Studies in Applied Mathematics}, 55:213--229, 1976.

\bibitem{ClKi93}
Ch. Claude, Yu.~S. Kishvar, O.~Kluth, and K.~H. Spatscheck.
\newblock Moving modes in localized nolnlinear lattices.
\newblock {\em Physical Review B}, 47(21):14228--14232, June 1993.

\bibitem{EiLo85}
J.~C. Eilbeck, P.~S. Lombdahl, and A.~C. Scott.
\newblock The discrete self-trapping equation.
\newblock {\em Phyisica}, 16 D:318--338, 1985.

\bibitem{ItIz93}
A.~R. Its, A.~G. Izergin, V.~E. Korepin, and N.~A. Slavnov.
\newblock Temperature correlations of quantum spins.
\newblock {\em Physical Review letters}, 70(11):1704--1706, March 1993.

\bibitem{VaGa87}
A.~A. Vakhnenko and Yu.~B. Gaididei.
\newblock On the motion of solitons in discrete molecular chains.
\newblock {\em Theoretical and Mathematical Physics}, 68(3):873--880, March
  1987.

\bibitem{AbCl91}
M.~J. Ablowitz and P.~A. Clarkson.
\newblock {\em Solitons Nonlinear Evolution Equations and Inverse Scattering}.
\newblock Number 149 in London Mathematical Society Lecture Note Series.
  Cambridge University Press, 1991.

\bibitem{Su94}
Y.~B. Suris.
\newblock A discrete-time {G}arnier system.
\newblock {\em Physics Letters. A}, 189(4):281--289, 1994.

\bibitem{Oh98}
Y.~Ohta.
\newblock Pfaffian solutions for coupled discrete nonlinear schr{\"o}dinger
  equation.
\newblock {\em Chaos, Solitons and Fractals}, to appear.
\newblock Proceedings of Brussels Meeting II: Integrability and Chaos in
  Discrete Systems (Brussels, 2-6 July 1997).

\end{thebibliography}

\begin{table}[p]
  \begin{center}
    $a_2 = 1$ \\
    \begin{tabular}{|c|c|c|c|c|}
      \hline
      $\bc_2^-$ & $|\bc_1^- \cdot \bc_1^+|$ & $|\bc_2^- \cdot \bc_2^+|$ &
      $\log_{10} \Delta_1$ & $\log_{10} \Delta_2$\\
      \hline
      $(0.0, 1.0)$   &  1.000  &  1.000  &  -7.57  &  -7.57 \\
      $(0.2, .98)$   &  0.575  &  0.575  &  -7.54  &  -7.54 \\
      $(0.4, .92)$   &  0.492  &  0.492  &  -7.52  &  -7.52 \\
      $(0.6, 0.8)$   &  0.624  &  0.624  &  -7.58  &  -7.58 \\
      $(0.8, 0.6)$   &  0.806  &  0.806  &  -7.56  &  -7.56 \\
      $(1.0, 0.0)$   &  1.000  &  1.000  &  -7.57  &  -7.57 \\
      \hline
    \end{tabular} \\
    \vspace{12pt}
    $a_2 = 3$ \\
    \begin{tabular}{|c|c|c|c|c|}
      \hline
      $\bc_2^-$ & $|\bc_1^- \cdot \bc_1^+|$ & $|\bc_2^- \cdot \bc_2^+|$ &
      $\log_{10} \Delta_1$ & $\log_{10} \Delta_2$\\
      \hline     
      $(0.0, 1.0)$   &  1.000  &  1.000  &  -8.92  &  -6.21 \\
      $(0.2, .98)$   &  0.900  &  0.999  &  -7.75  &  -6.12 \\
      $(0.4, .92)$   &  0.617  &  0.996  &  -7.41  &  -6.06 \\
      $(0.6, 0.8)$   &  0.204  &  0.993  &  -7.27  &  -6.07 \\
      $(0.8, 0.6)$   &  0.396  &  0.994  &  -7.37  &  -6.12 \\
      $(1.0, 0.0)$   &  1.000  &  1.000  &  -7.61  &  -6.25 \\
      \hline
    \end{tabular}
    \caption{
      Two-Soliton interaction for the symmetric system \eqref{sydvnls}.
      The soliton amplitude parameters are $a_1 = 1$ and $a_2 = 1$ or
      $a_2=3$ as it is given in each of the tables. For these values of
      $a_j$, the soliton width is comparable to the grid size and the solution
      is not ``close'' to the continuum limit.
      The soliton speed parameters, $b_1 = .1$ and $b_2 = - .1$, are such
      that the solitons move slowly relative to one another thereby
      increasing the strength of the interaction.
      The polarizations before interaction are $\bc_1^- = (1, 0)$ and
      $\bc_2^-$ as it is given in the tables.
      The polarizations after interaction are $\bc_1^+$ and
      $\bc_2^+$. $\Delta_j$ where $j=1,2$, is the difference, as given
      by eq. \eqref{error}, between the numerical
      solution at the final time and an exact solitary wave.
      The values $|\bc_j^- \cdot \bc_j^+| < 1$ indicate that the
      polarization vectors are shifted by the soliton interaction.
      The small errors, $\Delta_j$, show that the solitary waves interact
      nearly elastically with the measured deviation accounted for
      by error in the numerical time integration.
      }
    \label{table:twosym}
  \end{center}
\end{table}

\end{document}